\documentclass[12pt,preprint]{emulateapj}

\begin{document}

\shorttitle{Modeling Thermal Emission from Millisecond Pulsars}
\shortauthors{Bogdanov, Rybicki, \& Grindlay}

\title{Constraints on Neutron Star Properties from X-ray Observations \\ of Millisecond Pulsars}

\author{Slavko Bogdanov, George B. Rybicki, and Jonathan E. Grindlay} 
\affil{sbogdanov@cfa.harvard.edu, grybicki@cfa.harvard.edu,
jgrindlay@cfa.harvard.edu
\\ Harvard-Smithsonian Center for
Astrophysics, 60 Garden Street, Cambridge, MA 02138}

\begin{abstract}
We present a model of thermal X-ray emission from hot spots on the
surface of a rotating compact star with an unmagnetized light-element
atmosphere. An application to \textit{ROSAT}, \textit{Chandra}, and
\textit{XMM-Newton} X-ray observations of the nearest known
rotation-powered millisecond pulsar (MSP) PSR J0437--4715 reveals that
the thermal emission from this pulsar is fully consistent with such a
model, enabling constraints on important properties of the underlying
neutron star.  We confirm that the observed thermal X-ray pulsations
from J0437--4715 are incompatible with blackbody emission and require
the presence of an optically thick, light element (most likely
hydrogen) atmosphere on the neutron star surface. The morphology of
the X-ray pulse profile is consistent with a global dipole
configuration of the pulsar magnetic field but suggests an off-center
magnetic axis, with a displacement of $0.8-3$ km from the stellar
center. For an assumed mass of 1.4 M$_{\odot}$, the model restricts
the allowed stellar radii to $R=6.8-13.8$ km (90\% confidence) and
$R>6.7$ km (99.9\% confidence), which is consistent with standard NS
equations of state and rules out an ultracompact star smaller than its
photon sphere. Deeper spectroscopic and timing observations of this
and other nearby radio MSPs with current and future X-ray facilities
(\textit{Constellation-X} and \textit{XEUS}) can provide further
insight into the fundamental properties of neutron stars.
\end{abstract}


\section{Introduction}

After four decades of considerable observational and theoretical
investigation of neutron stars (NSs), very little is known about
several key properties of these objects. In particular, there is very
limited information concerning their magnetic field configuration,
surface properties, and structure and composition of their interiors.
Thermal radiation from the physical surface of a NS can potentially
serve as a useful tool in the study of these properties.  For
instance, a measure of the surface temperature can provide insight
into the structure of the stellar interior through comparison with
theoretical NS cooling models \citep[see, e.g.,][and references
therein]{Page98}, although it relies on an accurate age determination
which is not easily measurable for most systems. In addition, mapping
the temperature variations accross the surface may be used to deduce
the topology of the surface magnetic field of strongly magnetized
NSs. Surface emission can also be used to infer the compactness of the
star through measurement of the gravitational redshift $z_g$. In
practice, this endeavor has proven to be quite difficult due to the
absence of spectral lines \citep[see, however,][]{Cott06}. For a NS
radiating from the entire surface, the gravitational redshift can be
obtained by determining the effective emission radius, $R_{\rm
eff}$. However, this method assumes that the inferred radius is
exactly equal to the radius at infinity, $R^{\infty}=(1+z_g)R$, of the
star. As such it is subject to uncertainties in the the emission
properties of and exact temperature distribution across the stellar
surface, as well as the distance to the source.

A promising approach is to study systems in which the thermal
radiation is confined to a very small fraction of the stellar surface
($R_{\rm eff}\ll R$). Such emission geometry is observed in rotation-
and accretion-powered milisecond pulsars (MSPs) as well as old normal
pulsars. In these objects, modeling the spectrum and rotation-induced
variations of the observed flux can provide a measure of the
compactness of the NS \citep[see][]{Pavlov97,Zavlin98}.  Moreover, the
hot spots on the surface are believed to coincide with the polar caps,
where the magnetic field is anchored to the NS surface, thus, allowing
one to determine the geometry of the NS magnetic field. In addition,
the modulation of the observed flux caused by the changing projection
of the emission area permits study of the radiation pattern of the
emergent intensity (isotropic versus anisotropic).

Herein, we present model spectra and lightcurves of compact stars with
hot spots covered by a hydrogen atmosphere.  An application of our
model to archival \textit{ROSAT}, \textit{Chandra}, and
\textit{XMM-Newton} observations of the radio MSP J0437--0437 permits
us to gain valuable insight into several fundamental NS
parameters. This work represents an extension of similar studies by
\citet{Pavlov97} and \citet{Zavlin98}. The present paper is organized
in the following manner. In \S2 we outline the theoretical model,
while in \S3 we describe its practical application. We offer a
discussion in \S4 and give conclusions in \S5.

\section{THEORETICAL MODEL}

Models of rotating compact stars with hot spots have been presented in
a host of publications
\citep[e.g.,][]{Pech83,Ftac86,Riff88,Mill98,Bra00,Belo02,Pou03,Vii04,Cad06}. However,
with some notable exceptions
\citep[e.g.,][]{Zavlin95a,Pavlov97,Zane06}, these studies have assumed
the idealized case of blackbody emission, i.e. an emergent intensity
independent of angle. Here, we examine in-depth the observable
properties of compact stars with light-element atmospheres and
illustrate the key differences from blackbody emission. Below, we
outline the basic formalism describing the system geometry and photon
propagation.

\subsection{Geometry}

The basis of the model consists of a compact star of mass $M$, radius
$R$, spin period $P$, and two antipodal hot spots\footnote{In \S3 we
also consider the case when the hot spots are not antipodal.} with
$R_{\rm eff}\ll R$. We designate the hot spot that approaches closer
to the observer as the primary and define the closest approach as
rotational phase zero ($\phi=0$).  The magnetic axis is inclined at an
angle $\alpha$ with respect to the spin axis, while the spin axis is
at an angle $\zeta$ relative to the line of sight to a distant
observer. The position of the primary hot spot on the NS surface is
defined by the angle between the normal to the surface and the line of
sight:
\begin{equation}
\cos\psi(t)=\sin \alpha \sin \zeta \cos \phi (t) + \cos \alpha \cos \zeta
\end{equation}
An expression for the secondary hot spot is easily
obtained by substituting $\alpha \to \pi-\alpha$ and $\phi \to
\pi+\phi$. Note that $\alpha$ and $\zeta$ are interchangeable as they
figure in equation (1) in the same way.

In Schwarzschild geometry, due to the compact nature of the NS, the
gravitational field has a profound effect on the photons as they
propagate from the stellar surface to infinity.  In particular, the
energy of a photon is reduced by a factor of $1+z_g=(1-R_S/R)^{-1/2}$,
where $R_S=2GM/c^2$, as it escapes from the deep gravitational
potential \citep[see, e.g.,][]{MTW70}.  Furthermore, a photon emitted
at an angle $\theta>0$ with respect to the local radial direction
follows a curved trajectory and is observed at infinity at an angle
$\psi>\theta$. The relation between these two angles is given by
\citep{Pech83}:
\begin{equation}
\psi= \int_{R}^{\infty}\frac{{\rm d}r}{r^2}\left[\frac{1}{b^2}-\frac{1}{r^2}\left(1-\frac{R_S}{r}\right)\right]^{-1/2}
\end{equation}
where 
\begin{equation}
b=\frac{R}{\sqrt{1-R_S/R}}\sin\theta
\end{equation}
is the impact parameter at infinity of a photon emitted from radius
$R$ at an angle $\theta$.  In many practical situations, one can use a
more convenient approximate relation between $\psi$ and $\theta$
\citep{Zavlin95a,Belo02}
\begin{equation}
\cos\psi\approx\frac{\cos\theta-R_S/R}{1-R_S/R}
\end{equation}
which is valid for $R > 2R_S$ and is remarkably accurate (fractional
error of order a few percent).  One interesting consequence of the
bent photon trajectories is that most of the NS surface is
observable at any given time.  In flat spacetime, the visibility
condition is simply $\cos\psi=\cos\theta>0$, while in Schwarzschild
geometry a point on the neutron star surface is visible down to a
critical angle $\cos\psi_c$, corresponding to $b_{\rm max}\equiv
R^{\infty}=R/\sqrt{1-R_S/R}$, the apparent radius of the star.

\subsection{Doppler Effect and Time Delays}

For rotation- and accretion powered MSPs, the rapid
motion of the NS surface induces an appreciable Doppler effect,
parameterized through the familiar Doppler factor
\begin{equation}
\eta=\frac{1}{\gamma(1-v/c \cos \xi)}
\end{equation}
where $\gamma=1/\sqrt{1-(v/c)^2}$, $v=2\pi R/P(1-R_S/R)^{-1/2}
\sin\alpha$ is the velocity of the hot spot as measured in the
inertial frame of the hot spot, and $\xi$ is the angle between the
direction of the velocity vector and the line of sight. As shown by
\citet{Pou03} and \citet{Vii04}, this angle can be expressed in terms
of $\theta$, $\psi$, $\alpha$, and $\phi$ as
\begin{equation}
\cos\xi=-\frac{\sin\theta}{\sin\psi}\sin \alpha \sin \phi
\end{equation}
where the quantity $\sin\theta/\sin\psi$ can be approximated by its
asymptotic value for small angles $(1-R_S/R)^{1/2}$.

We note that the assumption of a spherical Kerr metric does not lead
to an appreciable difference in the shape of the lightcurves compared
to the Schwarzschild case \citep[as demonstrated by][]{Cad06}.  We
also emphasize that for $P\lesssim3$ ms deviations from spherical
symmetry such as rotation-induced oblateness becomes important.
\citet{Cad06} have found that in that regime neither the spherical
Schwarzschild nor spherical Kerr metrics provide an accurate
description of the properties of the rotating star. Therefore, the
validity of our model is limited to spin periods greater than $\sim$3
ms.

Photons emitted from the far side of the NS relative to the observer,
in addition to following a curved trajectory, have to travel a greater
distance compared to a photon emitted radially. The resulting time lag
of the photon as seen by a distant observer is given by the elliptical
integral \citep{Pech83}
\begin{equation}
\Delta t(b)=\frac{1}{c} \int_{R}^{\infty}\frac{{\rm d} r}{1-R_S/R} \Bigg\{\left[1-\frac{
b^2}{r^2}\left(1-\frac{R_S}{r}\right)\right]^{-1/2}-1\Bigg\}
\end{equation}
This time delay translates into a phase lag ($\Delta\phi$) of a photon
\begin{equation}
\Delta \phi =\frac{2 \pi}{P}\Delta t
\end{equation}
which yields the observed phase $\phi_{\rm obs} = \phi+\Delta \phi$
\citep{Vii04}.  For $R/R_S\approx2.5$, the maximum $\Delta t$,
corresponding to a photon with maximum impact parameter $b_{\rm
max}=(1+z_g)R$, is of order $60$ $\mu$s, or $\lesssim$6\% of the
pulsar period for $P>1$ ms. Therefore, propagation time differences
have a minor effect on the observed lightcurves even for the most
rapidly rotating neutron stars. Nonetheless, this effect has been
included in our model for completeness.

The observed flux per unit frequency from each spot is given by
\begin{equation}
F(\nu)=I(\nu){\rm d}\Omega
\end{equation}
where $I(\nu)$ is the intensity of the radiation as measured at infinity
and ${\rm d}\Omega$ is the apparent solid angle subtended by the hot
spot.  Transforming both quantities to the rest frame of the hot spot
yields
\begin{equation}
F(\nu)=(1-R_S/R)^{1/2}\eta^3 I'(\nu',\theta')\cos\theta'\frac{{\rm d} \cos\theta}{{\rm d} \cos\psi} \frac{{\rm d}S'}{D^2}
\end{equation}
where the primed quantities are measured in the NS surface rest frame
\citep{Pou03}, with $\cos\theta'=\eta\cos\theta$ and ${\rm
d}S\cos\theta={\rm d}S'\cos\theta'$.  $I'(\nu',\theta')$ is the
emergent intensity, ${\rm d}S'$ is the emission area and $D$ is the
distance.  The three Doppler factors arise from the transformation of
the intensity. A fourth factor is obtained upon integration of
equation (10) over a frequency interval since ${\rm
d}\nu=(1-R_S/R)^{1/2}\eta{\rm d}\nu'$.  The total observed flux for a
given rotational phase is found by relating $\phi$ and $\theta$ for
each hot spot through $\psi$ via equations (1) and (2), using the
corresponding $I'(\nu',\theta')$ in equation (10), and summing the
flux from the two hot spots.  Equation (10) allows one to generate a
phase-integrated spectrum by simply integrating over the whole
rotation period.  It can also be used to construct an arbitrary
emission region on the NS surface by considering multiple surface
elements.

%
%

\begin{figure}
\begin{center}
\includegraphics[width=0.47\textwidth]{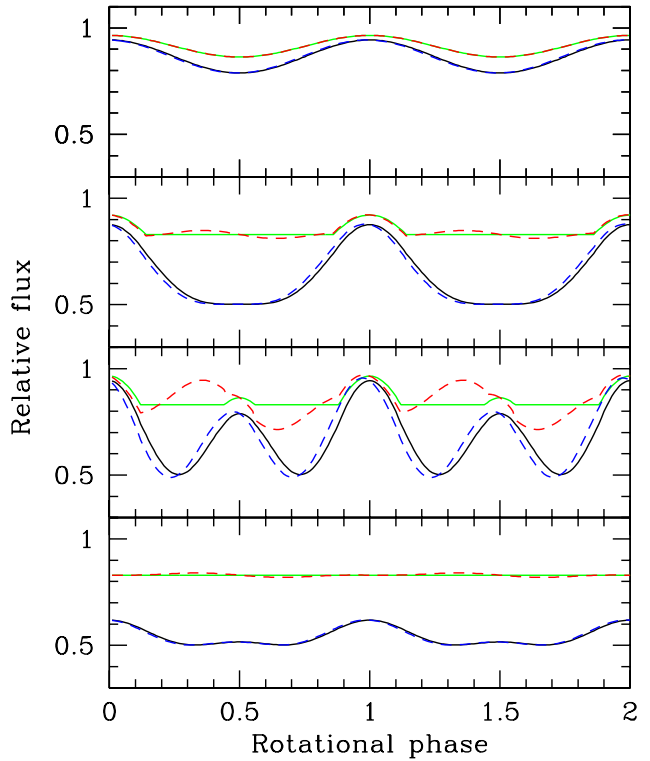}

\caption{Representative model lightcurves for a rotating $M=1.4$
M$_\odot$, $R=10$ km NS with two point-like antipodal hot spots. The
panels show classes I ($\alpha=10^{\circ}$, $\zeta=30^{\circ}$), II
($\alpha=30^{\circ}$, $\zeta=60^{\circ}$), III ($\alpha=60^{\circ}$,
$\zeta=80^{\circ}$), and IV ($\alpha=20^{\circ}$, $\zeta=80^{\circ}$)
as defined by Beloborodov (2002). The solid lines in each plot
correspond to a hydrogen atmosphere (\textit{black}) and blackbody
emission (\textit{grey}) with no Doppler effect included. The dashed
lines show the effect of Doppler boosting and aberration for $P=4$ ms.
All fluxes are normalized to the value for
$\alpha=\zeta=0$. Two rotational cycles are shown for clarity.}
\end{center}
\end{figure}

\subsection{Atmosphere Model}

To describe the emission properties of the stellar surface we use the
hydrogen atmosphere model from \citet[][]{McC04}\citep[see
also][]{Heinke06}\footnote{This model is virtually identical to that
presented by \citet{Zavlin96}.} having the following set of
assumptions. The atmosphere is static, in radiative equilibrium, and
composed purely of hydrogen. The scale heights in the atmosphere are
much smaller than the NS radius so a plane-parallel approximation is
valid.  The star is weakly magnetized ($B\ll10^{9}$ G), meaning that
the effects of the magnetic field on the opacity and equation of state
of the atmosphere can be ignored \citep[see e.g.,][for a treatment of
magnetized atmospheres]{Shib92,Zavlin95b}. The opacity within the
atmosphere is due to thermal free-free absorption plus Thomson
scattering in the unpolarized, isotropic approximation.  For the
temperatures of interest ($\sim$$10^6$ K) we expect complete
ionization implying that bound-free and bound-bound transitions are
unimportant. Finally, this model is valid for temperatures below
$T_{\rm eff}\approx 3 \times 10^6$ K, where the effects of
Comptonization within the atmosphere are neglegible.

An important characteristic of light-element NS atmosphere models is
that the peak emission occurs at higher energies than a blackbody (BB
hereafter) for the same effective temperature
\citep{Rom87,Zavlin96}. Furthermore, the emission pattern of the
atmosphere is inherently anisotropic with intensity decreasing as a
function of angle with respect to the normal, resulting in a
limb-darkening effect \citep[cf Fig.~7 of][]{Zavlin96}.  This implies
that although the emission spectrum of such an atmosphere is
qualitatively similar to the case of a BB, the observed
rotation-induced variation of the hot spot flux will be profoundly
different. Figure 1 shows a sample of lightcurves, generated using
equation (10) for a range of $\alpha$ and $\zeta$ for both BB and H
atmosphere models, assuming $M=1.4$ M$_{\odot}$ and $R=10$ km. The
most notable observable property of the atmosphere lightcurves is the
substantially larger ``depth'' (i.e. pulsed fraction) relative to the
BB case. Effectively, the limb darkening of the atmosphere acts to
partially negate the effect of light bending, which tends to reduce
the degree of modulation. Note, however, that even in the case of
point-like hot spots, the modulations are still quite broad. In
addition, it is not possible to achieve arbitrarily large pulsed
fractions ($\sim$100\%) for any combination of $\alpha$ and
$\zeta$. These characteristics can be used to distinguish between
thermal and non-thermal pulsed emission in cases where the spectral
continuum alone does not provide sufficient information \citep[see
e.g.][for the case of PSR J0737--3039A]{Cha07}.  Limb darkening also
significantly diminishes the effect of Doppler boosting on the shape
of the pulse profile relative to the BB model, as evident in Figure
1. This is because, for a given hot spot, when the component of the
velocity vector along the line of sight is largest (corresponding to a
maximum Doppler boost) the hot spot provides a very small contribution
to the total flux.

Unlike a BB, the pulsed fraction of the atmosphere emission is
expected to vary depending on the choice of energy band due to the
energy-dependent optical depth of the hydrogen layer \citep{Zavlin96}.
This marked difference between the atmosphere and BB lightcurves
allows one to discriminate between the two emission models. For
phase-resolved spectra of an atmosphere we expect softening of the
emission at the pulse minima, due to the presence of a temperature
gradient within the atmosphere (with the coolest layer at the
surface). For many systems, however, only phase-integrated spectra are
available due to inadequate photon statistics and/or time
resolution. Such model X-ray spectra in which the geometry, strong
gravity, and rotation of the NS have been taken into account are shown
in Figure 2.  It is apparent that in addition to a change in the total
flux, there is a significant shift in the peak energy of the spectrum.
For a BB spectrum no such variations are expected. Therefore, even in
the case of phase-integrated spectra it is important to take into
account the system geometry ($\alpha$ and $\zeta$) to ensure reliable
measurements of the temperature and effective area of the hot spots.

%
%
\begin{figure}[!t]
\begin{center}
\includegraphics[width=0.45\textwidth,angle=270]{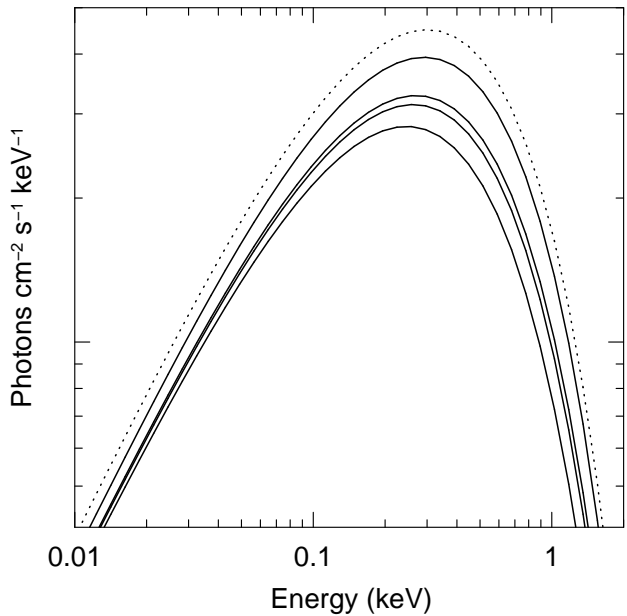}
\caption{Phase-integrated synthetic H atmosphere hot spot emission
spectra for $M=1.4$ M$_{\odot}$, $R=10$ km, and $T_{\rm eff}=2\times
10^6$ K. The solid lines correspond to classes I, III, II, and IV,
from top to bottom, respectively, with values of $\alpha$ and $\zeta$
as in Figure 1. The dotted line is for a star with one hot spot always
face-on ($\alpha=\zeta=0$). For all spectra the same arbitrary $R_{\rm
eff}$ is assumed. Note the shifts in peak energy.}
\end{center}
\end{figure}

%
%
\begin{figure}[t]
\begin{center}
\includegraphics[width=0.47\textwidth]{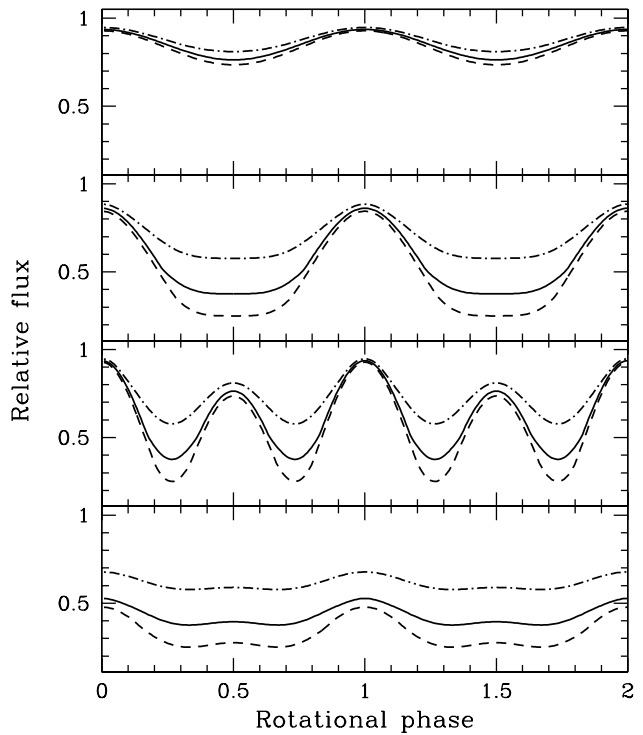}
\caption{Model H atmosphere lightcurves for different stellar radii
for an assumed $1.4$ M$_\odot$ star. The curves correspond to $R=9$ km
(\textit{dot-dashed}), 12 km (\textit{solid}), and 16 km
(\textit{dashed}). The values of $\alpha$ and $\zeta$ for each panel
are the same as in Fig. 1. Two rotational cycles are shown for
clarity.}
\vspace{-0.5cm}
\end{center}
\end{figure}

\section{APPLICATION}

Close examination of equations (2) and (10) reveals that the flux
observed at infinity from the hot spots depends on several important
NS parameters. Figure 3 illustrates the effect of the stellar
compactness on the pulse profile of a rotating NS with two antipodal
hot spots covered with a H atmosphere. It is apparent that a small
increase in $M/R$ results in a substantial decrease in the pulsed
fraction. This is a consequence of the strong sensitivity of light
bending on the compactness of the star. Thus, as shown by
\citet{Pavlov97} and \citet{Zavlin98}, modeling of the spectra and
pulse profile may, in principle, allow stringent constraints on $M$
and $R$. Note that in the case of atmosphere models, $M$ and $R$ also
figure implicitly in the intensity $I'(\nu',\theta')$ through the
surface gravity $g_s=(1-R_S/R)^{-1/2}M/R^2$. Although this dependence
is much weaker than that through equation (2), it suggests that $M$
and $R$ should be treated as separate parameters. The spectra and
lightcurves also depend strongly on the angles $\alpha$ and $\zeta$
through equations (1) and (2).  In rotation-powered pulsars, $\alpha$
corresponds to the pulsar obliquity, i.e. the angle between the spin
and magnetic axes. Knowledge of this angle has important implications
in the study of NS magnetic fields. Hence, an application of our model
to X-ray spectroscopic and timing observations may yield useful
constraints on $M/R$, $\alpha$, and $\zeta$.

To allow direct comparison of our model with observational data we
have generated a grid of H atmosphere emergent intensities,
$I'(\nu',\theta')$, for a range of effective temperatures $T_{\rm
eff}$, $g_s$ (implying a range of $M$ and $R$), and $\theta$. This
grid allows us to generate a spectrum and pulse profile for any
desired values of $M$, $R$, $\alpha$, and $\zeta$.  To minimize the
number of free parameters, we fix $M$ to 1.4 M$_{\odot}$ and vary only
$R$. For the purposes of spectral analysis, we have incorporated our
model into
XSPEC\footnote{http://heasarc.gsfc.nasa.gov/docs/xanadu/xspec/}
12.3.0. The fits to the data are performed as in \citet{Pavlov97} and
\citet{Zavlin98}. We first fit the spectrum for assumed values of $R$,
$\alpha$, and $\zeta$. The resulting best fit $T_{\rm eff}$, $R_{\rm
eff}$ (the effective radius of each hot spot), and $N_{\rm H}$ (the
integrated H column density along the line of sight) are then used to
generate the corresponding pulse profiles.  After convolving with the
appropriate detector response and accounting for background noise, the
model lightcurves are binned to the same number of phase bins as the
data, which allows a calculation of $\chi^2$
\begin{equation}
\chi^2 = \sum_{i=1}^{K}\frac{(N_{o,i}-N_{m,i})^2}{N_{o,i}+N_{b,i}}
\end{equation}
where $N_{o,i}$, $N_{m,i}$, and $N_{b,i}$ are the observed, model, and
background counts in the $i$th phase bin, respectively, and $K$ is the
total number of bins.  This procedure is repeated for a range of $R$,
$\alpha$, $\zeta$, and $N_{\rm H}$ to map the multidimensional
$\chi^2$ surface and derive confidence levels for these parameters.

\begin{deluxetable}{llccc}
\tabletypesize{\small} 
\tablecolumns{5} 
\tablewidth{0pc}
\tablecaption{X-ray Observations of PSR J0437--4715}
\tablehead{ \colhead{Observatory} & \colhead{Detector} & {Energy band} & {Exposure} & \colhead{Note\tablenotemark{a}} \\
\colhead{ } & \colhead{ } & \colhead{(keV)} & \colhead{(ks)} & \colhead{ }}

\startdata

\textit{ROSAT}      &  PSPC      &  0.1--2  & 9.8  &  S+T   \\
\textit{Chandra}    &  ACIS-S    &  0.3--8  & 25.7  &  S      \\
                    &  HRC-S     &  0.1--10 &  19.6 &  T      \\
\textit{XMM-Newton} &  EPIC-MOS &  0.3--10 & 68.3  &  S    \\
		    &  EPIC-pn   &  0.3--10 & 67.2  &  S+T   
\enddata 

\tablenotetext{a}{S=spectroscopy, T=timing.}
\tablerefs{Zavlin \& Pavlov 1998; Zavlin et al. 2002; Zavlin 2006.}

\end{deluxetable}

\subsection{The Radio Millisecond Pulsar PSR J0437--4715}

Recent X-ray observations have revealed that the emission from most
rotation-powered MSPs \citep{Beck02,Grind02,Zavlin06,Bog06a} is of
predominantly thermal nature. This radiation is believed to originate
from the magnetic polar caps of the pulsar that are heated by a
back-flow of high-energy particles from the magnetosphere \citep[see
e.g.,][] {Hard02a}.  Given that the progenitors of radio MSPs undergo
an extended period of accretion during the low-mass X-ray binary phase
\citep{Alp82,Bha91} it is reasonable to expect these stars to have
accumulated a substantial surface layer of gas. Due to the immense
surface gravity the accreted material quickly stratifies, with the
lightest element remaining on top and dominating the observed
emission. Therefore, the existence of a thick light element atmosphere
at the surface of MSPs is highly likely \citep{Zavlin95a}. Note that
for the quality of the available spectral data of MSPs, at the
temperatures relevant for these sources a He atmosphere is virtually
indistinguishable from a H one, although based on the standard
formation scenario of MSPs the existence of H at the surface is more
probable. Also, \citet{Raj96} and \citet{Zavlin02} have shown that a
heavy element (Fe) atmosphere does not provide a good description of
MSP spectra.

%
%
\begin{figure}[t!]
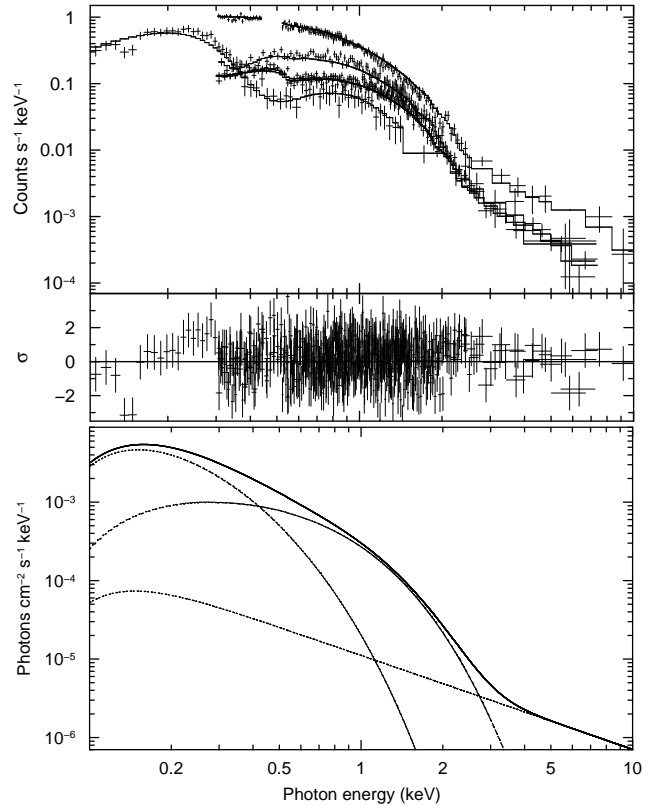

\begin{center}
\includegraphics[angle=270,width=0.47\textwidth]{f4a.ps}\\
\includegraphics[angle=270,width=0.47\textwidth]{f4b.ps}
\caption{(\textit{Upper panels}) \textit{ROSAT} PSPC, \textit{Chandra}
ACIS-S, and \textit{XMM-Newton} EPIC-MOS1/2 and EPIC-pn spectra of PSR
J0437--4715. The lines show a representative best fit of our model to
the data. The bottom panel shows the best fit
residuals. \textit{(Lower panel)} Representative best-fit model
spectrum consisting of two hydrogen atmosphere and one PL components.}
\end{center}
\end{figure}

Of the $\sim$40 X-ray detected radio MSPs, at present, only PSR
J0437--4715 \citep{John93}, the nearest and brightest, has X-ray data
of sufficient quality to permit detailed modeling of its spectrum and
pulse profile. \citet{Zavlin02} and \citet{Zavlin06} have shown that
the spectrum of this pulsar is well described by a multi-component
model consisting of two dominant hydrogen atmosphere thermal
components with temperatures $T_{\rm{1}}=(2.1-1.5) \times 10^6$ K and
$T_{\rm{2}}=(0.52-0.54) \times 10^6$ K and effective radii
$R_{1}\approx0.34$ km and $R_{2}\approx2.0$ km, for assumed $M=1.4$
M$_{\odot}$, $R=10$ km, $\alpha=45^{\circ}$ and
$\zeta=45^{\circ}$. The presence of two temperatures can be attributed
to non-uniform heating of the polar caps by the return flow of
magnetospheric particles \citep[see][]{Hard02a, Zhang03}.

A faint power-law (PL) tail is also observed from this MSP at energies
$\gtrsim$2 keV. The nature of this PL emission is ambiguous as it may
arise either due to magnetospheric emission processes or weak
Comptonization of the thermal radiation by the tenuous magnetospheric
plasma (Bogdanov et al. 2006b). A crucial difference between the two
interpretations is that in the former the PL component extends to
energies well below the soft X-ray band, while in the latter the PL is
only present at higher energies ($\gtrsim$2 keV). Moreover, the
magnetospheric emission PL is inconsistent with the FUV flux from this
MSP \citep{Kar04} unless its spectral photon index is
$\Gamma\lesssim1.6$.

Previosly, \citet{Pavlov97} and \citet{Zavlin98} have applied a
two-temperature hydrogen atmosphere polar cap model to spectroscopic
and timing \textit{ROSAT} PSPC observations of PSR J0437--4715 in an
effort to constrain $M/R$ of this NS. In this analysis, a centered
dipole field was considered. Based on this assumption a limit
of $R>8.8$ km for $M=1.4$ M$_{\odot}$ was obtained.  As deeper
observations of this MSP have become available since, it is worth
revisiting this object to determine if better constraints on $M/R$ can
be obtained.

We have retrieved archival \textit{ROSAT}, \textit{Chandra}, and
\textit{XMM-Newton} spectral and timing observations of this MSP to
investigate its X-ray emission. These observations are summarized in
Table 1.  It is important to note that the values for the effective
temperatures and radii derived by \citet{Zavlin02} and
\citet{Zavlin06} are by no means unique as they are sensitive to the
initial choice of $M$, $R$, and (as evident in Fig. 2) $\alpha$ and
$\zeta$. We have carried out a series of spectral fits for $M=1.4$,
$R=8-16$ km, $\alpha=0^{\circ}-90^{\circ}$,
$\zeta=0^{\circ}-90^{\circ}$\footnote{For either $\alpha$ or $\zeta$,
the range $90^{\circ}-180^{\circ}$ becomes equivalent to
$90^{\circ}-0^{\circ}$ if we reassign the secondary hot spot to be the
primary and apply the transformation $\phi \to \phi + \pi$. Therefore,
it is only necessary to consider the latter range.}, and $N_{\rm
H}=(1-3)\times10^{19}$ cm$^{-2}$. For this set of parameters, we find
$T_1=1.43-1.85$ MK, $R_1=0.04-0.40$ km, $T_2=0.4-0.56$ MK, and
$R_2=1.8-4.9$ km, with $\chi_{\nu}^2=1.3$ for 402 degrees of freedom.
The ranges quoted give the $\pm$1$\sigma$ bounds for one interesting
parameter.  Given the known cross-calibration uncertainties of the
various detectors, the minimum value of $\chi_{\nu}^2$ corresponds to
an acceptable fit and indicates a systematic uncertainty at the level
of $\sim$5--10\%. These results are generally in good agreement with
those derived in previous studies.  We have performed the spectral
fits for two cases: PL emission with $\Gamma\lesssim1.6$ and no PL
emission below 2 keV as an approximation to a Comptonization model.
In both cases, we find that such a component contributes with
$\lesssim$5\% to the total photon flux below 2 keV. Furthermore,
although the statistics at high energies are quite limited,
\textit{XMM-Newton} EPIC-pn data above 3 keV suggest that the PL
does not show sharp pulsations \citep[cf Fig. 3 of][]{Bog06b} as
observed in the more luminous MSPs, PSRs B1937+21 \citep{Nic04} and
B1821--24 \citep{Rut04}. This implies that the observed pulse shape is
not significantly affected by the PL emission. Therefore, in our
lightcurve analysis we will include this component as a constant (DC)
contribution to the pulsed flux.

\subsection{Fits to the Pulse Profile of PSR J0437--4715}

Using the derived range of temperatures and radii we proceed to
generate synthetic pulse profiles and compare them to those of
J0437--4715. When folded at the spin period ($P=5.76$ ms), the X-ray
emission from this MSP exhibits a single broad pulse, with pulsed
fraction 30-40\%, depending on the choice of energy band
\citep[see][]{Zavlin06}, consistent with thermal radiation.  A visual
comparison with the lightcurves in Figure 1 shows that the X-ray
pulsations of this MSP (Fig. 5) most closely resemble those of classes
I and II. However, it is immediately apparent that the pulses of
J0437--4715 are asymmetric, characterized by a faster decrease than
increase in flux.  This is clearly inconsistent with the antipodal hot
spot model.  In addition, at the spin period of this MSP the Doppler
effect is fairly weak so it cannot reproduce the degree of asymmetry
observed. Indeed, an acceptable fit could not be obtained for any
combination of $R$, $\alpha$, and $\zeta$. The discrepancy can be
resolved if the two hot spots are not diametrically opposite, i.e. if
the magnetic axis is off-center. Therefore, in our model we introduce
two additional free parameters, $\Delta\phi$ and $\Delta\alpha$,
offsets in $\phi$ and $\alpha$, respectively, of the secondary hot
spot.  These correspond to displacements on the NS surface in
longitude and latitude, respectively, from the antipodal position. The
net offset on the stellar surface of the secondary hot spot from the
antipodal position is
\begin{equation}
\Delta s = R \cos ^{-1}[-\sin\alpha\sin(\alpha+\Delta\alpha)-\cos\alpha\cos(\alpha+\Delta\alpha)\cos\Delta\phi]
\end{equation}
In an offset dipole configuration $\alpha$ is no longer the angle
between the spin and magnetic axes and the dipole axis does not run
through the center of the star. The total displacement (i.e. impact
parameter) of the magnetic axis from the center is then
\begin{equation}
\Delta x = R\sin\left(\frac{\Delta s}{2R}\right)
\end{equation}
If we introduce an offset dipole, we find that a phase lag of $\Delta
\phi \sim -20^{\circ}$, implying $\Delta x \sim 1$ km, is able to account for the asymmetry in the
pulse profile.

In the formal fits to the X-ray pulse profiles of PSR J0437--4715 we
consider nine free parameters: $T_1$, $R_1$, $T_2$, $R_2$, $R$,
$\alpha$, $\zeta$, $\Delta\alpha$, and $\Delta\phi$, while keeping
$N_{\rm H}$ fixed at $2\times10^{19}$ cm$^{-2}$ and $M$ at 1.4
M$_{\odot}$. In addition, since the relative phase between the various
observations cannot be determined due to the inadequate absolute
timing precision of \textit{ROSAT} and \textit{XMM-Newton}, we allow
the $\phi$ to vary independently for each dataset.  The model was
fitted individually to \textit{ROSAT} PSPC (0.1--2.4 keV),
\textit{Chandra} HRC-S (0.1--10 keV), and \textit{XMM-Newton} (0.3--2
keV) timing data. For the latter, the availability of spectral
information and the sufficient photon statistics permit us to consider
two energy bands: 0.3--0.7\footnote{For this energy range, the
0.44--0.51 keV band was excluded to eliminate an instrumental feature
specific to the fast timing mode of the EPIC-pn detector.} and 0.7--2
keV, where the cool and hot thermal component dominate,
respectively. Since the effects of the system geometry and strong
gravity are purely achromatic, the availability of multiple energy
bands is very useful as it enables a test of the validity of the
atmosphere model, which exhibits characteristic energy dependent
effects (as described in \S2.2).  Due to the 10-fold better photon
statistics of the \textit{XMM-Newton} dataset, compared to the
\textit{ROSAT} PSPC and \textit{Chandra} HRC-S data, it provides the
best constraints on the desired parameters. We have considered the
other data, nonetheless, as they provide a useful consistency check of
the model.  The best-fit parameters and confidence intervals were
determined by manually searching the $\chi^2$ hyperspace for the
minimum. In principle, an additional strong constraint can be imposed
by requiring that the derived temperatures and radii from these fits
be within 3$\sigma$ of those obtained from the spectral
analysis. Unfortunately, due to the limited photon statistics as well
as the relatively large cross-calibration uncertainties of the
detectors, which are clearly evident in Figure 4, the 3$\sigma$ range
of derived temperatures ($T_1=1.35-1.95$ MK and $T_2=0.39-0.60$ MK) is
not particularly constraining.

%
%
\begin{figure}
\begin{center}
\includegraphics[angle=270,width=0.47\textwidth]{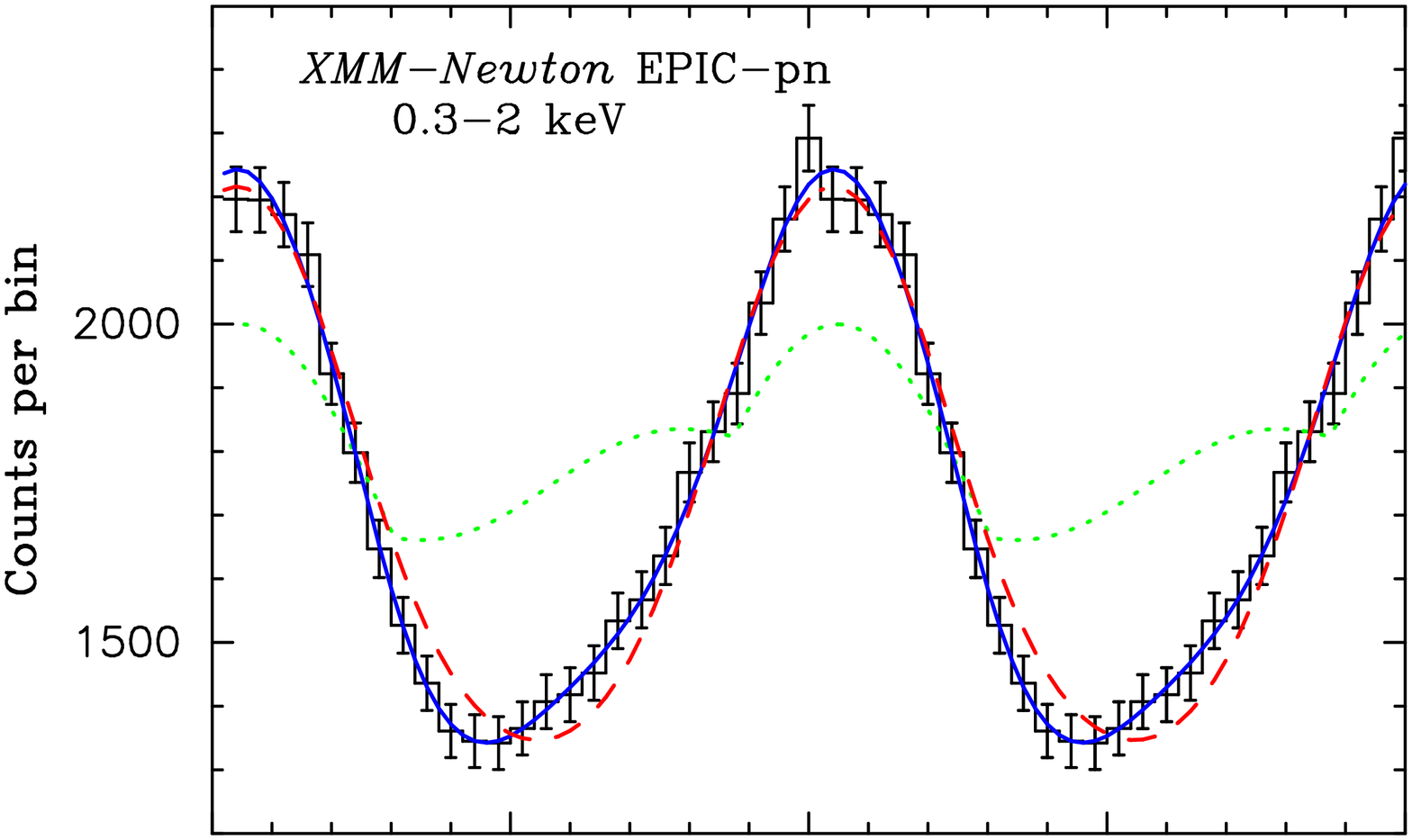}\\
\includegraphics[angle=270,width=0.47\textwidth]{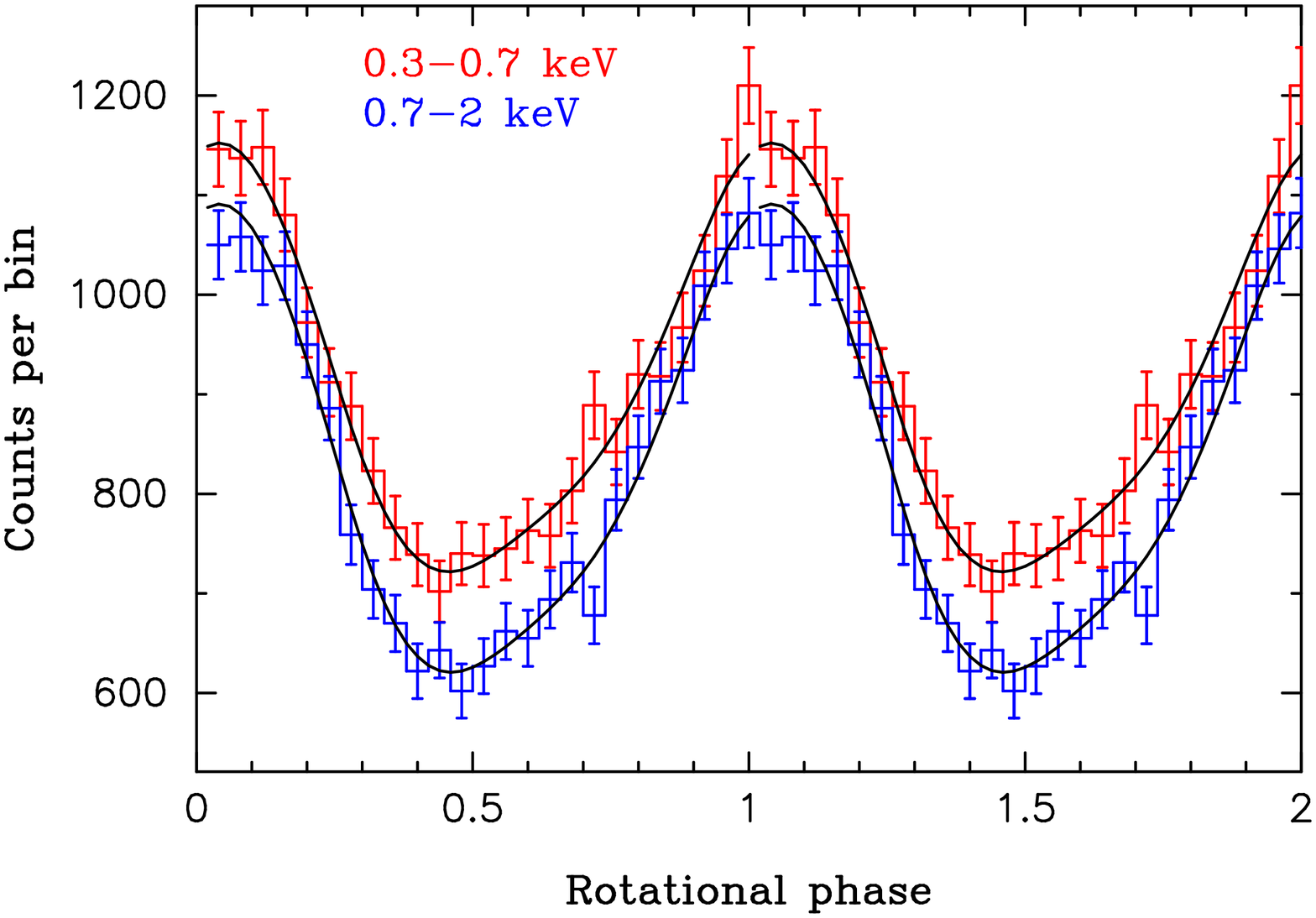}
\caption{(\textit{Top}) \textit{XMM-Newton} EPIC-pn 0.3--2 keV pulse
profile of PSR J0437--4715. The solid line shows the best fit curve
for an off-center dipole field and a H atmosphere with $R=9.6$ km,
$\alpha=30^{\circ}$, $\zeta=42^{\circ}$,
$\Delta\alpha=-12^{\circ}$, $\Delta\phi=-20^{\circ}$, $T_1=2.02$ MK,
$R_1=0.2$ km, $T_2=0.48$ MK, $R_2=3.5$ km, and $N_H=2\times10^{19}$
cm$^{-2}$. The dotted line is of a blackbody that best fits the
spectrum of J0437--4715 for the same assumed compactness and
geometry. The dashed line corresponds to the best fit H atmosphere
model for a centered dipole.  (\textit{Bottom}) Best fits to the
\textit{XMM-Newton} EPIC-pn pulse profiles in the 0.3--0.7 keV and
0.7--2 keV bands (upper and lower curve, respectively) assuming an
off-center dipole. The choice of phase 0 is arbitrary.}
\end{center}
\end{figure}

%
%
\begin{figure}[!t]
\begin{center}
\includegraphics[angle=270, width=0.47\textwidth]{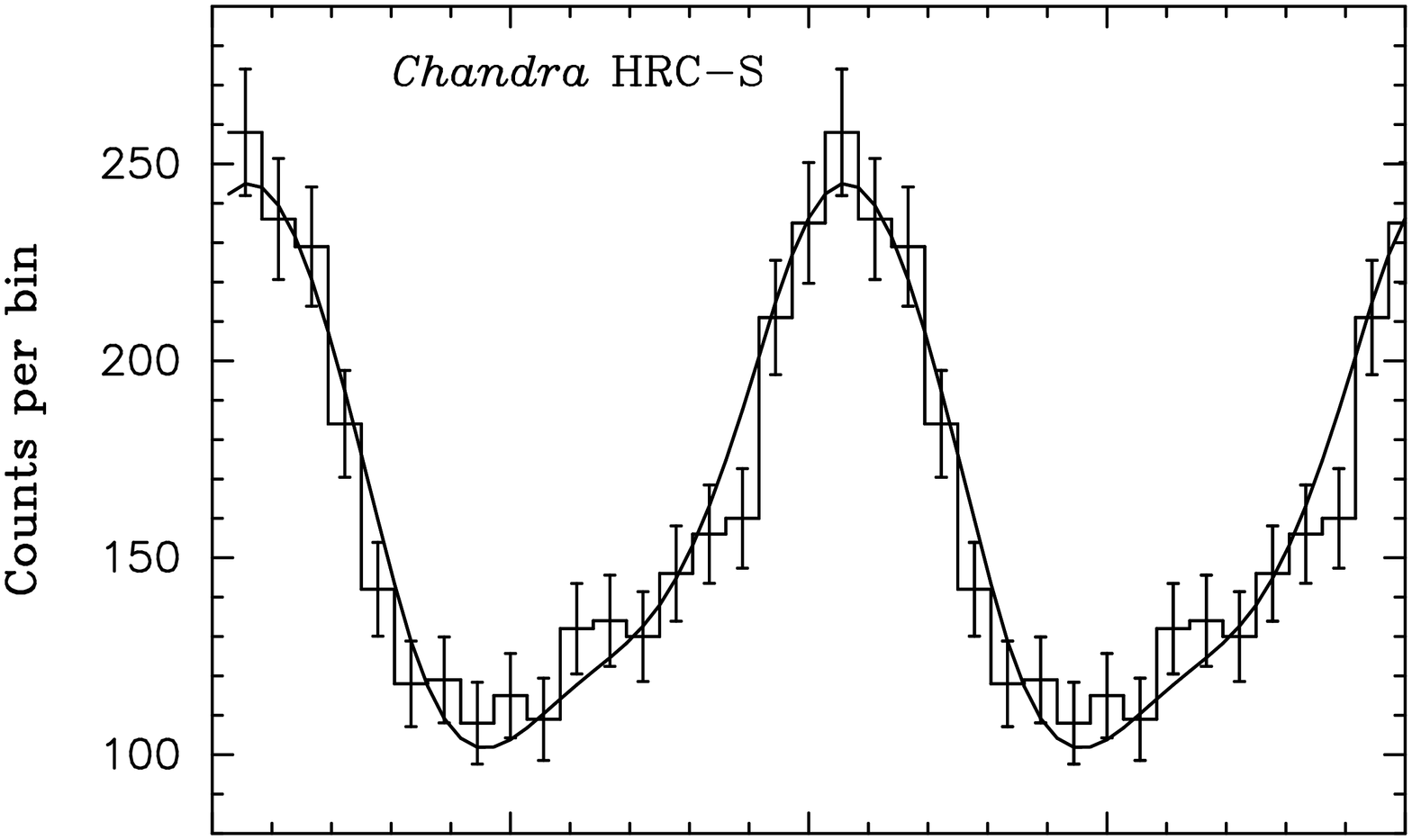}\\
\includegraphics[angle=270, width=0.47\textwidth]{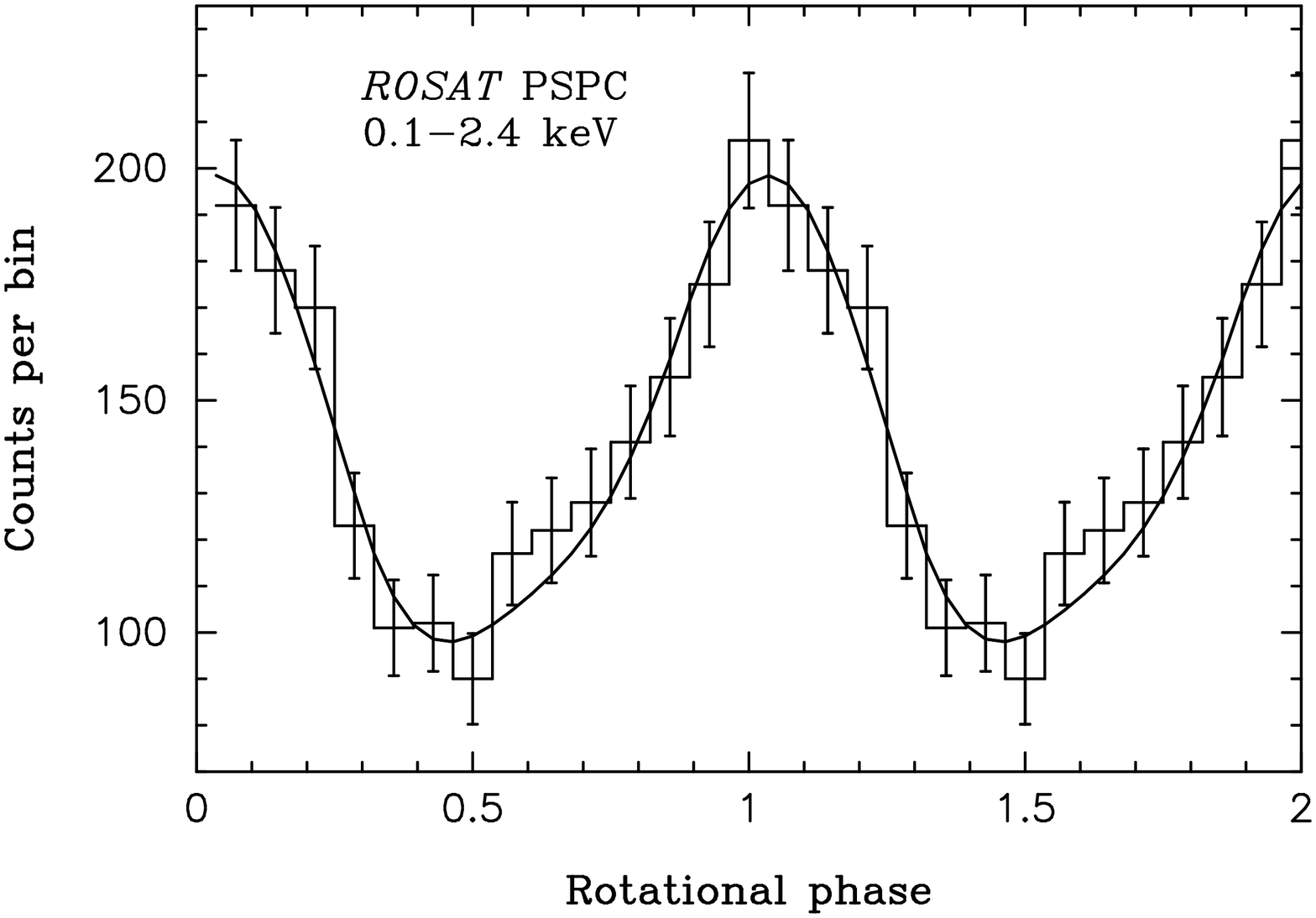}
\caption{\textit{Chandra} HRC-S (\textit{top}), and \textit{ROSAT}
PSPC (\textit{bottom}) pulse profiles of J0437--4715. The solid lines
show the best fit model to the \textit{XMM-Newton} EPIC-pn data. The
flux normalization $R_{\rm eff}^2/D^2$ has been adjusted to correct
for the discrepancies in the absolute calibration of the different
detectors (see text for details).}
\end{center}
\end{figure}

The atmosphere model is able to reproduce the shape of the pulse
profile and variation in pulsed fraction with energy remarkably well,
as is evident in Figures 5 and 6.  As a consequence of the numerous
free parameters and the complex covariances that exist between them,
acceptable values of $\chi^2$ were obtained for a relatively wide
range of parameter values. Fortunately, the strong correlations can be
weakened if we fix the angle $\zeta$ to its most likely value.  Radio
timing observations indicate that the orbital inclination of the
J0437--4715 binary system is $i=42^{\circ}$ \citep{van01, Hot06}. It
is highly probable that the spin axis of the MSP is closely aligned
with the orbital anguar momentum vector as a direct result of the
spin-up process during the LMXB phase. If this is indeed the case, the
angle between the spin axis and the line of sight must be
$\zeta\approx42^{\circ}$. This constraint significantly constricts the
allowed parameter space and thus improves the accuracy with which the
other parameters can be determined. In particular, for the fit to the
\textit{XMM-Newton} EPIC-pn pulse profiles we find the $\pm$1$\sigma$
ranges to be: $R=6.9-10.6$ km, $\alpha=25^{\circ}-90^{\circ}$, $\Delta
\alpha=-50^{\circ}-20^{\circ}$, $\Delta\phi=-(23^{\circ}-14^{\circ})$,
$T_1=1.4-1.85$ MK, $T_2=0.4-0.54$ MK, $R_1=0.1-0.36$ km, and
$R_2=2.0-3.5$ km.  All ranges quoted represent 1$\sigma$ confidence
intervals. Varying $N_{\rm H}$ over the range $(1-3)\times10^{19}$
cm$^{-2}$ does not cause significant changes in the derived results
since the interstellar extinction towards this MSP is neglegible above
$\sim$0.3 keV. The derived values of $\Delta\phi$ and $\Delta \alpha$
translate into an offset of the magnetic axis from the center of the
star of $\Delta x = 0.8-3$ km (1$\sigma$). For $R$ we are also able to
derive 90\% and 99.9\% limits of $6.8-13.9$ km and $>$6.7 km,
respectively (see Fig. 7); similar limits are obtained even without
the constraint on $\zeta$.  The fits to the \textit{ROSAT} PSPC and
\textit{Chandra} HRC-S pulse profiles are generally consistent with
these results if we account for the systematic uncertainties
encountered in the spectral fits.  We note in passing that the best
fit parameters for PSR J0437--4715 derived by \citet{Zavlin98} using
the \textit{ROSAT} PSPC as well as their assumption of a centered
dipole do not yield acceptable fits to the \textit{Chandra} HRC-S and
\textit{XMM-Newton} EPIC-pn pulse profiles, even if we account for
cross-calibration uncertainties.

We have found that $\alpha$ and $\Delta\alpha$ are the two parameters
that are most strongly covariant with $R$ (see
Fig. 8). $\Delta\alpha$, in particular, closely mimics the effects of
$M/R$ on the observed pulse profiles (Fig. 3), thus weakening the
constraint on $R$. If $\alpha$ and $\Delta\alpha$ for this MSP can be
determined by independent means (e.g., from radio polarization
measurements), the accuracy of the measurement of $R$ can be greatly
improved.

We have conducted the entire fitting procedure outlined above for a BB
emission model as well. We find that the thermal portion of the X-ray
spectrum is well described by a two-temperature BB model. This is
expected, given the close qualitative similarities between the BB and
H atmosphere spectra. However, a BB model is incapable of reproducing
the pulse shape and pulsed fraction for any $R$.  Furthermore, a BB
cannot account for the observed changes in the pulsed fraction as a
function of photon energy (see Fig. 5). Therefore, we confirm the
finding of \citet{Pavlov97} and \citet{Zavlin98} that \textit{a
blackbody is not a valid description of the surface of J0437--4715,
implying that an optically thick, light-element atmosphere must be
present on the stellar surface}.

%
%
\begin{figure}[!t]
\begin{center}
\includegraphics[width=0.48\textwidth]{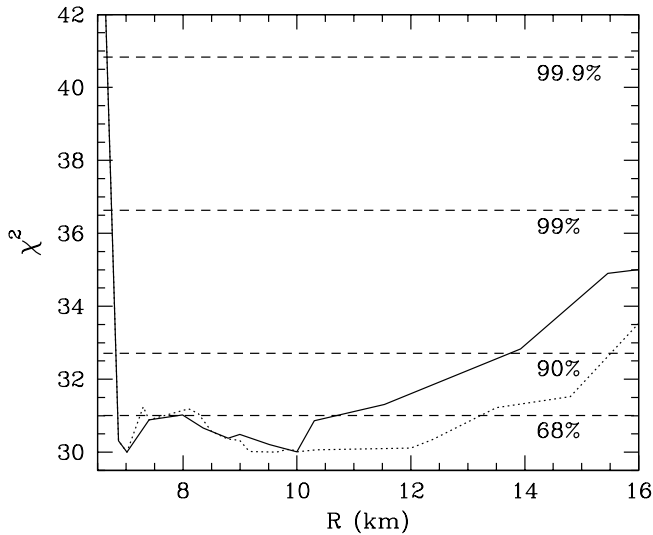}
\caption{The effect of the stellar radius on the $\chi^2$ statistic of
the fit to the \textit{XMM-Newton} EPIC-pn pulse profile of PSR
J0437--4715 for $M=1.4$ M$_{\odot}$. The solid line corresponds to a
fixed value of $\zeta=42^{\circ}$, while the dotted line is for no
constraint on $\zeta$. The dashed lines show the 68\%, 90\%, 99\%, and
99.9\% confidence levels for one interesting parameter.}
\end{center}
\end{figure}

%
%
\begin{figure}[!t]
\begin{center}
\includegraphics[width=0.48\textwidth]{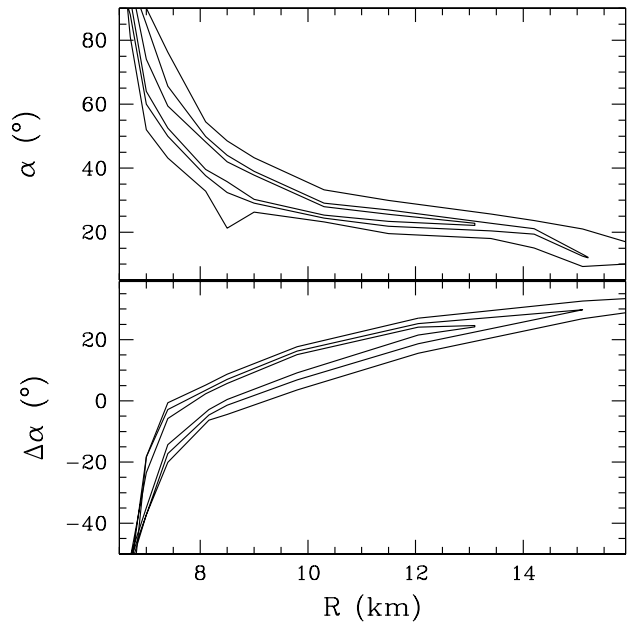}

\caption{68\%, 90\%, and 99.9\% confidence contour plots showing the
correlation between the stellar radius $R$ and $\alpha$ (top) and
$\Delta\alpha$ (bottom) for $M=1.4$ $M_{\odot}$ in the fit to the
\textit{XMM-Newton} EPIC-pn pulse profile of PSR J0437--4715.}
\end{center}
\end{figure}

\section{DISCUSSION}

The principal motivation of our analysis is to extract information
regarding the compactness of the NS, magnetic field configuration,
and radiative properties of the stellar surface.  It is important to
realize, however, that the method of inferring these properties is
inherently model dependent. Thus, it is necessary to examine the
validity of the approximations and assumptions we have made as well as
any potential bias that they may introduce in the derived
results. This is especially critical for determination of $M/R$ if
this approach is to be used to place stringent constraints on the NS
equation of state (EOS).

One of the underlying assumptions of the atmosphere model is that the
source of heat is below the atmospheric layer. At first, this seems to
be at odds with the prediction of pulsar electrodynamic models that
the MSP polar caps are heated by a return current of ultra-relativistic
electrons/positrons from the magnetosphere above the polar caps.
However, the penetration depth of the impinging particles is
significantly larger than the characteristic depth of the hydrogen
atmosphere. In particular, the stopping column of a relativistic
electron in hydrogen, assuming energy loss via bremsstrahlung and a
negligible magnetic field, is $y\sim60$ g cm$^{-2}$ \citep[][and
references therein]{Tsai74}. For comparison, in the case of a hydrogen
atmosphere with $\log T_{\rm eff}\sim5.6-6.4$ the characteristic depth
is $\sim$0.01--1 g cm$^{-2}$ \citep[cf Fig. 2
of][]{Zavlin96}. Therefore, the bulk of the energy of the returning
particles is deposited well below the depth at which the atmosphere
becomes optically thick.

For PSR J0437--4715, the relatively long spin period ($P=5.76$ ms)
suggest that the rapid rotation has a negligible effect on the shape
of the star. Hence, our assumption of spherical Schwarzschild geometry
is appropriate. Also, based on the pulsar period ($P=5.76$ ms) and
acceleration-corrected period derivative ($1.86 \times 10^{-20}$ s
s$^{-1}$) we find that the magnetic field strength at the surface of
the MSP polar caps is $B_{\rm
surf}\propto(P\dot{P})^{1/2}\approx3.3\times10^{8}$ G so the
assumption of an unmagnetized atmosphere is correct \citep{Zavlin95a}.

In our analysis of PSR J0437--4715 we have ignored any thermal
emission outside of the polar cap regions. \textit{Hubble Space
Telescope} FUV observations of this MSP \citep{Kar04}
indicate that the rest of the star is at a temperature $T_{\rm
eff}\sim10^5$ K. Given that this emission peaks in the UV, its
contribution to the X-ray flux is insignificant.

Another simplifying assumption we have introduced is that of
point-like emission regions.  To determine the effect of this
approximation, we have compared model pulse profiles of point-like and
extended circular emission areas with $R_{\rm eff}\sim 2-4$ km,
comparable to those inferred for the cooler emission component of the
spectrum of PSR J0437--4715. We find that the maximum difference
between the two cases, occurring at $\phi\approx 0$ and
$\phi\approx0.5$, is only $\sim$1\%, which is significantly lower than
the statistical uncertainty of the presently available timing data.
Therefore, the hot spots are truly point-like so this approximation is
more than adequate. Note that for X-ray timing data with better photon
statistics, the exact size and shape of the emission regions may
become important and may, in principle, be measurable quantities.

We have also assumed that the two thermal components occupy the same
($\phi$, $\alpha$) location on the NS surface.  The lack of any
appreciable phase shift between the cool (0.3--0.7 keV) and hot
(0.7--2 keV) components, as determined from the highest quality
(\textit{XMM-Newton} EPIC-pn) data, implies that the two emission
regions are effectively concentric. The detection of any such phase
difference would provide interesting information regarding the
temperature distribution across the polar caps.

It is important to emphasize that the possible presence of faint PL
emission below $\sim$2 keV in the spectrum of PSR J0437--4715 does not
prohibit reliable constraints on the properties of the underlying
NS. This component only introduces a small increase in the uncertainty
of our measurement. With future deep phase-resolved spectroscopic
observations the true nature of the PL can be determined and the
temporal properties of this component can be properly modeled,
resulting in improved constraints on the desired NS properties.

Along with insufficient photon statistics, one of the main factors
limiting the accuracy of our measurement of $M/R$ is the uncertainty
in the calibration of the X-ray detectors used in this analysis.  As
noted in \S3, this problem precludes a precise determination of
$T_{1}$ and $T_{2}$ from the spectral fits, which can provide much
tighter constraints on $M/R$ than obtained in our analysis. Note that
the uncertainty in detector performance has lesser impact on the fits
to the pulse profiles, which are integrated over a relatively wide
energy band so that deviations in the spectral response tend to
average out. Moreover, this uncertainty mostly effects the flux
normalization $R_{\rm eff}^2/D^2$ of the pulse profile and has little
bearing on the validity of our measurement of $R$ and the conclusions
regarding the presence of an atmosphere on the NS surface and magnetic
field geometry.  Improvement in the detector calibration will enable
much tighter constraints on $M/R$.

\section{CONCLUSIONS}

We have presented a model of thermal emission from rotating NSs
covered with an unmagnetized hydrogen atmosphere, applicable to
systems for which the radiation is localized in small regions on the
surface. Such a configuration is expected in MSPs. Indeed, the model
lightcurves (Figs. 1 and 3) are qualitatively similar to the observed
thermal pulse profiles of nearby MSPs in the field of the Galaxy
\citep{Beck02,Zavlin06}. In addition, the broad modulations and
moderate pulsed fractions ($\lesssim$50\%) of the lightcurves produced
by our model are in full agreement with the limits obtained for the
thermal MSPs in 47 Tuc \citep{Cam07}.

An application to the nearest radio MSP J0437--4715 shows that the
spectral and temporal properties of its X-ray emission are easily
reproduced by our model, without the need to resort to contrived
assumptions regarding the properties of the NS atmosphere
\citep[composition and thickness, see, e.g.,][]{Ho07,Mori07} and the
temperature distribution across the stellar surface. This excellent
agreement allows valuable insight into important NS properties that
are not measurable by other means.  Specifically, we confirm that the
radiative properties of the NS surface are fully consistent with a
light element atmosphere model and are poorly reproduced by a
blackbody.  This provides compelling evidence that the surface of this
MSP is indeed covered by a gaseous layer of hydrogen.  Note that this
result does not necessarily apply to all varieties of NS systems given
that MSPs are formed via a special evolutionary channel, involving
prolonged accretion of gas from a close stellar companion.  The pulse
profile of J0437--4715 suggests that the global magnetic field
configuration closely resembles a dipole, albeit one significantly
offset from the stellar center.  It is unclear whether this
displacement is primordial, i.e., the result of the core-collapse that
formed the NS, the result of magnetic field burial during the LMXB
phase or a consequence of magnetic field evolution and
migration as proposed by \citet{Rud91}.  We note that small scale
deviations from a pure dipole, with characteristic scales $\ll$R, may
still exist near the NS surface.  Ultra-deep X-ray spectroscopic and
timing observations may, in principle, reveal such ``microstructure'',
which could be manifested in the form of temperature variations across
the face of the polar caps.

Perhaps the most important practical application of our model is a
constraint on the compactness of the NS, which has important
implications for determination of the true NS EOS. In particular, the
restriction of $R>6.7$ km (at 99.9\% confidence) for a $M=1.4$
M$_{\odot}$ star implies that PSR J0437-4715 is not ultra-compact,
i.e. not smaller than its photon sphere. For this and other binary
MSPs a very accurate, independent measurement of the NS mass is
possible using radio timing observations.  Thus, in principle, a
unique determination of the stellar radius can be achieved.  Note that
for J0437--4715, the two mass measurements of $M=1.58\pm0.18$
\citep{van01} and $M=1.3\pm0.2$ M$_{\odot}$ (Hotan et al. 2006)
currently available do not provide very tight constraints on the
allowed masses.

Despite the inherent model dependence of this approach and the
relatively large number of free parameters we are able to extract
valuable information regarding the poorly understood properties of
pulsars and neutron stars, in general. Morover, as discussed in \S4,
the conclusions drawn from our analysis are fairly robust in the sense
that they are weakly sensitive to the underlying assumptions and
approximations.  Future deep observations of PSR J0437--4715 and other
MSPs can provide even more stringent constraints on fundamental NS
parameters.  One close MSP ($D\approx300$ pc), PSR J0030+0451
\citep{Beck02,Lom06}, exhibits a double peaked X-ray pulse profile,
indicating a viewing angle and pulsar obliquity substantially
different from those of J0437--4715. As such, this pulsar is ideal for
an independent verification of this method. An upcoming deep
\textit{XMM-Newton} observation will allow a detailed study of PSR
J0030+0451.

MSPs are much better suited for this analysis than other NS systems
such as X-ray binaries, isolated NSs, and normal pulsars. In the
latter objects there are numerous complications arising due to the
effects of the strong magnetic field on the emergent thermal
radiation, the unknown temperature distribution across the surface,
severe reprocessing of the thermal radiation by the magnetosphere, and
the uncertain emission altitude above the NS surface \citep[e.g. in
X-ray bursts,][]{Cott06}.  Their low magnetic fields, point-like
emission regions, and non-variable, dominant thermal emission make
MSPs suitable laboratories for tests of fundamental NS physics and
constraints on their EOS.

 We would like to thank Ramesh Narayan and Bryan
Gaensler for numerous insightful discussions. We also thank the
anonymous referee for many useful comments that helped improve the
manuscript. This work was supported in part by NASA grant AR6-7010X.
The research presented here has made use of the NASA Astrophysics Data
System (ADS).


\begin{thebibliography}{}

\bibitem[Alpar et al.(1982)]{Alp82} Alpar, M. A., Cheng, A. F., Ruderman, M. A., \& Shaham, J. 1982, Nature, 300, 728

\bibitem[Becker \& Achenbach(2002)]{Beck02} Becker, W. \& Achenbach, B. 2002, Proceedings of the 270. WE-Heraeus Seminar on Neutron Stars, Pulsars, and Supernova Remnants, ed. W. Becker, H. Lech, \& J. Tr\"umper, p.64

\bibitem[Beloborodov(2002)]{Belo02} Beloborodov, A. M. 2002, \apj, 566, L85

\bibitem[Bhattacharya \& van den Heuvel(1991)]{Bha91} Bhattacharya, D. \& van den Heuvel, E. P. J. 1991, Phys. Rep., 203, 1


\bibitem[Bogdanov et al.(2006a)]{Bog06a} Bogdanov, S., Grindlay, J. E., Heinke, C. O., Camilo, F., Freire, P. C. C, \& Becker, W. 2006a, \apj, 646, 1104

\bibitem[Bogdanov et al.(2006b)]{Bog06b} Bogdanov, S., Grindlay, J. E., \& Rybicki, G. B. 2006b, \apj, 648, L55

\bibitem[Braje et al.(2000)]{Bra00} Braje, T. M., Romani, R. W., \& Rauch, K. P. 2000, \apj, 531, 447

\bibitem[Cadeau et al.(2007)]{Cad06} Cadeau, C., Morsink., S. M., Leahy, D., \& Campbell, S. S. 2007, ApJ, 654, 458

\bibitem[Cameron et al.(2007)]{Cam07} Cameron, P. B., Rutledge, R. E., Camilo, F., Bildsten, L., Ransom, S. M., \& Kulkarni, S. R. 2007, \apj, in press 

\bibitem[Chatterjee et al.(2007)]{Cha07} Chatterjee, S., Gaensler,
B. M., Melatos, A., Brisken, W. F., \& Stappers, B. W. 2007, \apj, in
press (astro-ph/0703181)

\bibitem[Cottam et al.(2006)]{Cott06} Cottam, J., Paerels, F., \& Mendez, M. 2006, Nature, 420, 51

\bibitem[Ftaclas et al.(1986)]{Ftac86} Ftaclas, C., Kearney, M. W., \& Pechenick, K. 1986, \apj, 300, 203

\bibitem[Grindlay et al.(2002)]{Grind02} Grindlay, J. E., Camilo, F.,
Heinke, C.  O., Edmonds, P. D., Cohn, H., \& Lugger, P. 2002, \apj,
581, 470

\bibitem[Harding \& Muslimov(2002)]{Hard02a} Harding, A. K. \&
Muslimov, A. G. 2 002, \apj, 568, 862

\bibitem[Heinke et al.(2006)]{Heinke06} Heinke, C. O., Rybicki, G. B., Narayan, R., \& Grindlay, J. E. 2006, \apj, 644, 1090

\bibitem[Ho et al.(2007)]{Ho07} Ho, W. C. G., Kaplan, D. L., Chang. P., van Adelsberg, M., \& Potekhin, A. Y. 2007, MNRAS, 375, 821

\bibitem[Hotan et al.(2006)]{Hot06} Hotan, A. W., Bailes, M., \& Ord, S. M. 2006, MNRAS, 369, 1502

\bibitem[Johnston et al.(1993)]{John93} Johnston, S., Lorimer, D. R.,
Harrison, P. A., Bailes, M., Lyne, A. G., Bell, J. F., Kaspi, V. M.,
Manchester, R. N., D'Amico, N., \& Nicastro, L. 1993, Nature, 361, 613

\bibitem[Kargaltsev et al.(2004)]{Kar04} Kargaltsev, O. Y., Pavlov,
G. G., \& Romani, R. W. 2004, \apj, 602, 327

\bibitem[Lattimer \& Prakash(2001)]{Latt01} Lattimer, J. M. \&
Prakash, M. 2001, \apj, 550, 426

\bibitem[Lommen et al.(2006)]{Lom06} Lommen, A. N., Kipphorn, R. A., Nice, D. J., Splaver, E. M., Stairs, I. H., \& Backer, D. C. 2006, \apj, 642, 1012

\bibitem[McClintock et al.(2004)]{McC04} McClintock, J. E., Narayan,
R., \& Rybicki, G. B. 2004, \apj, 615, 402

\bibitem[Miller \& Lamb(1998)]{Mill98} Miller, M. C. \& Lamb, F. K. 1998,
\apj, 499, L37

\bibitem[Misner, Thorne, \& Wheeler(1970)]{MTW70} Misner, C. W,
Thorne, K. S., \& Wheeler, J. A. 1970, Gravitation (New York:
Freeman)

\bibitem[Mori \& Ho(2007)]{Mori07} Mori, K. \& Ho, W. C. G. 2007,
MNRAS, 377, 905

\bibitem[Nicastro et al.(2004)]{Nic04} Nicastro, L., Cusumano, G., L\"ohmer, O., Kramer, M., Kuiper, L., Hermsen, W., Mineo, T., \& Becker, W. 2004, A\&A, 413, 1065

\bibitem[Page(1998)]{Page98} Page, D. 1998, The Many Faces of Neutron
Stars, ed. R. Buccheri, J. van Paradijs, M. A. Alpar, p. 539

\bibitem[Pavlov \& Zavlin(1997)]{Pavlov97} Pavlov, G. G. \& Zavlin,
V. E. 1997, \apj, 490, L91

\bibitem[Pechenick et al.(1983)]{Pech83}
Pechenick, K. R., Ftaclas, C., \& Cohen, J. M. 1983, \apj, 274, 846

\bibitem[Poutanen \& Gierli\'nski(2003)]{Pou03} Poutanen, J. \& Gierli\'nski, M. 2003, MNRAS, 343, 1301

\bibitem[Rajagopal \& Romani(1996)]{Raj96} Rajagopal, M. \& Romani, R.~W. 1996, \apj, 461, 327

\bibitem[Riffert \& M\'esz\'aros(1988)]{Riff88} Riffert, H. \& M\'esz\'aros, P. 1988, \apj, 325, 207

\bibitem[Romani(1987)]{Rom87} Romani, R. W. 1987, \apj, 313, 718

\bibitem[Ruderman(1991)]{Rud91} Ruderman, M. 1991, \apj, 366, 261

\bibitem[Rutledge et al.(2004)]{Rut04} Rutledge, R. E., Fox, D. W., Kulkarni, S. R., Jacoby, B. A., Cognard, I., Backer, D. C., \& Murray, S. S. 2004, \apj, 613, 522

\bibitem[Shibanov et al.(1992)]{Shib92} Shibanov, Iu. A., Zavlin, V. E., Pavlov, G. G., \& Ventura, J. 1992, A\&A, 266, 313

\bibitem[Tsai(1974)]{Tsai74} Tsai, Y.-S. 1974, Rev.~Mod.~Phys, 46, 815

\bibitem[van Straten et al.(2001)]{van01} van Straten, W., Bailes, M.,
Britton, M., Kulkarni, S. R, Anderson, S. B., Manchester, R. N., \&
Sarkissian, J. 2001, \nat, 412, 158

\bibitem[Viironen \& Poutanen(2004)]{Vii04} Viironen, K. \& Poutanen, J. 2004, A\&A, 426, 985

\bibitem[Zane \& Turolla(2006)]{Zane06} Zane, S. \& Turolla, R. 2006, MNRAS, 366, 727

\bibitem[Zavlin et al.(1995a)]{Zavlin95a} Zavlin, V. E., Shibanov, Y. A., \& Pavlov, G. G. 1995a, Astron. Lett., 21, 149

\bibitem[Zavlin et al.(1995b)]{Zavlin95b} Zavlin, V. E., Pavlov, G. G., Shibanov, Y. A., \& Ventura, J. 1995b, A\&A, 297, 441

\bibitem[Zavlin et al.(1996)]{Zavlin96}
Zavlin, V. E.  , Pavlov, G. G., \& Shibanov, Yu. A., 1996, A\&A, 315,
141

\bibitem[Zavlin \& Pavlov(1998)]{Zavlin98} Zavlin, V. E. \& Pavlov, G. G. 1998, A\&A, 329, 583


\bibitem[Zavlin et al.(2002)]{Zavlin02} Zavlin, V. E., Pavlov, G. G.,
Sanwal, D.  , Manchester, R. N., Tr\"umper, J., Halpern, J. P., \&
Becker, W. 2002, \apj, 56 9, 894

\bibitem[Zavlin(2006)]{Zavlin06} Zavlin, V. E. 2006, \apj, 638, 951

\bibitem[Zhang \& Cheng(2003)]{Zhang03} Zhang, L. \& Cheng, K. S. 2003, A\&A,398, 639


\end{thebibliography}
\end{document}